\documentstyle[aps,amssymb]{revtex}

\newcommand{\beq}{\begin{equation}}
\newcommand{\eeq}{\end{equation}}
\newcommand{\beqar}{\begin{eqnarray}}
\newcommand{\eeqar}{\end{eqnarray}}

\begin{document}
\title{Toward a Theory of Marginally Efficient Markets\footnote{Based on presentations at the 
{\it econophysics} meetings in Rome, March 1998 and Palermo Sept.1998. To Appear in Palermo Proceedings (1999), Ed. R. N. Mantegna}}
\author{Yi-Cheng Zhang}
\address{Institut de Physique Th\'eorique,
Universit\'e de Fribourg, P\'erolles, Fribourg CH-1700, 
Switzerland}

\maketitle
\abstract{Empirical evidence suggests that even the most competitive markets
are not strictly efficient. Price histories can be used to predict near future returns with
a probability better than random chance. Many markets can be considered as {\it favorable games}, in the sense that there is a small probabilistic edge that smart speculators can exploit. We propose to identify this probability using conditional
entropy concept. A perfect random walk has this entropy maximized, and departure from the maximal
value represents a price history's predictability. We propose that market participants should be divided
into two categories: producers and speculators. The former provides the negative entropy into the price, upon which the latter  
feed. We show that the residual negative entropy can never be arbitraged away: infinite arbitrage capital is needed to make
the price a perfect random walk.}
\vskip 1cm

\section{Introduction}

One of the fundamental pillars of Modern Economics is the Efficient Market Hypothesis (EMH). Its essential meaning is that
if the market price were predictable, then these opportunities would be exploited to make a gain so that such opportunities
would disappear in a competitive and efficient market. Indeed, this proposition is very plausible and has a long history.
Bachelier, almost a century ago\cite{Bachelier}, has had the vision that the market prices should behave like a random walk.
Contemporary economists has gone great length to formalise the random walk concept, reaching the hign-water mark when Samuelson mathematically "proved" that
properly anticipated prices are random walk \cite{proof}. Empirical observations, however, are much less convincing.
Many authors have indeed found that market prices may contain some profit pockets. The proponents of EMH generally label
such observations as "anomalies" without significance. 

On the other hand, if the competitive markets are indeed so efficient as the proponents of EMH would like to be, then we 
are facing the enormous paradox: there is a huge industry (Wall Street!)
trying hard to anticipate market moves using essentially past information. These institutions can hardly be accounted for in the orthodox economics.
The fact that people can  cling to EMH so long is that there  is  so far no rival alternative convincing enough to oust the elegant, simple assumption.
After all, prices in competitive markets have to be rather efficient.
In this work we show that, while the basic arguments of the random walk are generally valid, i.e. market prices are not
{\it very} predictable. It is the small inefficiency margin which contains all the interesting
dynamics.

What is wrong in the naive EMH is that it implies that if there are arbitrage opportunities, they would disappear
instantly upon speculators' action. We emphasize that arbitrage opportunities in general are represented by probabilities.
Though they are favorable in probabilistic sense, they are not riskless. To profit from such opportunities speculators
would need large capital and bear certain risk. Thus this favorable probability is the speculators' edge. Upon increased participation of speculators, this marginal probability would shrink, but never disappear.
This is so because with the diminished marginal probability, it is even harder to make a profit. Still larger capital
is needed and risk incurred is also larger. To make this favorable marginal probability disappear, infinite capital is needed and
the return per capital invested would diminish to zero.

Therefore, it is this marginal probability which keeps the market competitive and dynamic, such that it is 
attractive to all participants. Speculators work harder to outguess the market, if they make profit by exploiting their
probabilistic edge, the market will be more efficient, but never be so in the absolute sense. So a competitive market can keep
its marginal probability low thanks to the fierce competition of the participants.

We propose that the alternative of Marginally Efficient Markets (MEM) to replace the sterile EMH. Instead regarding
the anomalies as mere annoyance, we show in the following sections that it is in this marginal region many interesting things
take place.

\section {Market Inefficiency and Probabilistic Edge of Speculators}      
 
Modern version of EMH is formulated in three versions: strong form, semi-strong form, and weak form. For a review we refer 
readers to a recent comprehensive volume \cite{CLM}. Here we show a few examples suggesting that even the weak form does not hold,
with varying degrees of violation. The weak form of EMH states that using the past price {\it alone}, you cannot predict
the future price movements. Strong and semi-strong versions allow other informations besides the price history itself.
Fama \cite{Fama} was the first to systematically measure possible violation of the weak form of EMH more than thirty years ago.
His conclusion is that, though some inefficiencies are detected, probably it is so insignificant that it is unlikely to yield
a profit exploiting such inefficiencies. His results gave the benefit of doubt to EMH and it is widely hailed as evidence
that EMH is after all a valid concept.

 As the first example we consider the daily price (close) of NYSE Composite Index, which cover 400 largest US stocks. The duration
is over 32 years from 1966.
The choice of this set of data is its 
availability on the web, free of charge (www.nyse.com). The period (1966-1998) covers much of the happenstance of the current
economic era. The simplest measure of the price predictability is the conditional probability.
Denote the index price value in time by $x(t)$, $t$ being discrete (days). From the time series 
we construct the price increments $\Delta x(t)=x(t)-x(t-1)$, which is the daily price variation. Let us further simplify
to consider only signs of the daily variations, thus the price price history is reduced to a binary sequence of $\pm 1$.
To evaluate the conditional probability, we consider for a given sequence ${\bf i}$ of $M$ bits, what is the probability to 
have a $+1$ (or $-1$) to follow. If $+1$ and $-1$ are both as likely with an ${\bf i}$ given, then we can conclude that the price
is indeed a random walk, as EMH proponents would like to believe. On the other hand, if $+1$ and $-1$ probabilities distinctly
differ and consistently, then this market is not efficient and strategies can be designed to gain a profit, with a favorable
probability. 
In Fig.1 we show the conditional probabilities with $M=3$ e.g. $p(+-+|-)$, i.e. we want to see whether we can predict 
the next day price movement
with the knowledge of the price history limited to 3 previous consecutive daily variations. For $M=3$ there are 8 combinations, in Fig.1 
we plot only the occurrence to $+1$ since $p({\bf i} |+)+p({\bf i} |-)=1$ for all ${\bf i}$. For instance consider only the events of past three
daily variations are $+-+$, then $+$ is more likely to appear than $-$, with the differential probabilities
about 65\% and 35\%, respectively. In principle, market participants with daily intervention capabilities can profit from this
inefficiency. Whenever they see a $+-+$, they can bet that market would go up and buy right after the observation $+-+$ is made.
This is not risk free arbitrage opportunity in the usual sense, but it does represent a game with favorable chances. 

However, one should be warned that it is not simply the conditional probability alone which represents the speculator's favorable
edge. Market conditions do change over time, a speculator may observe some repeated patterns during one period then would like to
draw some useful conclusions for the next period. Only the conditional probability consistent over some time can be of use to make a profit.
Let us examine the conditional probabilities over different periods in Fig.2. The original 32 years data is broken into
4 segments with 8 years each. We see that there is considerable overlap between two consecutive periods. If one speculator
design a method using previous 8 years observation, then he can use it for the next 8 years, albeit his edge is only the overlapping
part of the two probabilities. For shorter periods, e.g. from year to year, the overlap is even more pronounced. We may conclude that market's behavior
is slowly varying, much like glassy systems in physics. In Fig.2 we see also that the favorable edge diminishes over time.
One should recall that a lot of events have happened during the period, including computer trading, and most importantly, the introduction
of the futures market---making transactions of stocks much easier. However, potential newcomer
speculators should not be disappointed,
for the favorable edge does not seem to disappear, in the next few sections we shall explain why. 

In the above discussion transaction costs are not considered. Indeed, most 
recent arguments from the proponents of EMH try to argue away such inefficiencies with transaction costs. Since NYSE index 
comprises many stocks, it is not easy to evaluate the transaction costs. For this purpose we turn to another type of markets:
Foreign Currency Exchange (Forex) and commodities markets. In Fig.3 are plotted 2 major cross Forex rates and the price of silver during the year of 1996. Forex is an active, competitive market. It is
said that daily turnover is larger than all the equity turnovers combined. Most of the volume can be attributed to
speculation and hedging activities. These data are records of every 30 minutes, i.e. so called high frequency data, are sold by
{\sl Olsen \& Associates} (www.olsen.ch). We plot the conditional probabilities for the three prices.
There are two ways to study the data, one is to study them using fixed intervals, like for the NYSE data which is daily; another
is to let the data run until a given threshold value is surpassed. Take US dollar vs Swiss franc as example. Let us use the 
threshold value 0.01 or 100 points (A cross rate is quoted to the fourth significant decimal place, the smallest unit is called
a point or a pip). With variable intervals but fixed differences we want to make sure that fluctuations are significant 
enough to warrant attention. Transaction costs in Forex market have a clearer structure: even amateur speculators can expect
to pay only 3 to 10 points on a round-turn, buy a currency and sell it again. In Fig.3 we see that three prices 
give similar conditional probabilities. For instance upon observing a $+-$ sequence one can bet that a $+$ would follow
with about 70 percent probability, i.e. 7 out of 10 guesses can be right. Taking into account the overlapping
parts of the conditional probability if we divide the one year data in two halves, this marginal probability is slightly reduced, but nevertheless about
2/3. Out of three trades two come out right, you end up with about 100 points. Paying out 
transaction costs on all three round-turn
trades costing 30 points, one still ends up with a profit of 70 points. Recall that Forex trading is highly leveraged, a speculator
 can trade sizes many times of his own capital. So these tiny points can add up to huge profit.

Of course the above examples are not meant to be a "proof" that the market in question is not efficient. Indeed, some data do appear
to be rather efficient. The point is that one does not have to look for very hard to find violations of EMH. Let us quote a few other examples {\it en passant}, Dollar-Yen data do not present much interesting
patterns using the above method, so are Canadian and Australien currencies vs USD. Even the above three prices can have "poor"  and "good"   
years, for the data in the first half of 1998 similar analysis like that in Fig.4 the patterns are less pronounced.
But in no case the qualitative patterns changes, e.g. $p(+-|+)$ is never smaller than 1/2 over long time periods.
The lesson we draw here is that even the weakest form (i.e. transaction costs included) of EMH is violated, making the general mathematical proof irrelevant.
In the next section we examine the question: why is the market inefficient despite of diligent work (pressumed) of the
majority of the speculators? We shall see that even with the best possible methods, it is impossible to arbitrage away the
probabilistic edge.

In the above discussion we do not specify exactly how to actually exploit the probabilistical
edges. In reality market participants have all sorts of strategies, operating at different frequencies, and more importantly, information beyond the prices alone. For a simplified model of
market dynamics, called "Minority Games", competing and interacting population of speculators is
studied in details\cite{czs}.  

\section{Market Participants, Market Dynamics}

Many studies have been done on optimization of positive returns from markets, neatly summarized 
in portfolio theory\cite{elton}. It rarely mentioned in the literature that in order for
"smart" investors to make money someone must lose it in the first place. Standard portfolio theory deals with 
how to get 
best portfolios, without giving hint of the global picture where buyers and sellers are both considered. If an investor, convinced of the optimal portfolio theory, buys the stocks
by the prescribed proportions. One is tempted to ask the question: who are the sellers stupid enough
to give up such optimal choices? Portfolio theories and CAPM (Capital Asset Pricing Model\cite{capm}) are conspicuously silent on this point.
   A convenient answer often invoked implies that there are "smart" and "stupid" participants and markets make just redistribution of
wealth among these two groups. This in turn implies that markets are just a 
{\it zero-sum} game. However, even casual observations show that there are not that many stupid participants who 
consistently lose years after years, to "donate" money through the market's roulette. In general, it is fair to say that 
majority of market participants benefit from participating markets, even though money is conserved in the exchange. Then we are left with the puzzle: who injects the money? And why? 

We believe that there is a general explanation to this puzzle. To proceed we want to divide market participants into 
two categories: producers and speculators. The producers are defined as those who have to participate the markets for
their own need other than speculation. Producers are just a convenient name, they should include hedgers or even tourists
doing casual foreign currency trading no matter what today's rates are. Speculators include notably the so-called arbitrageurs,
big financial institutions, professional and amateur traders or even hedge funds (e.g. LTCM) themselves. In many cases the distinction
between the two categories is blurred: some participants at times behave like producers and 
at other times like speculators. In short speculators try to fine-tune their timing to take advantages from price movements. They do not have to use
the above method to detect market inefficiencies, there are plenty other parameters other than the price itself can yield
favorable predictions. However, in this work we limit ourselves to attack the weak form of EMH, i.e. only the information
from past prices is considered.

We shall see that it is the producers who pump the money into the markets which the speculators eventually take away. 
Is this fair? Why should the producers be so stupid to allow speculators take advantage of them? Should we label
the speculators the "parasites" of the markets, thus they ought be outlawed and eliminated? Or maybe they are useful at all? We shall see that the answer to these questions are surprisingly simple:
producers are indeed payers but $happy$ payers; speculators, on the other hand, though motivated solely by profit opportunities, perform a 
social function by making markets liquid. A liquid market is better for a producer since his need can be constantly met.

Why do we call producers "happy" payers? Let us digress a bit to consider the relationship between the insurer and the insured.
Take the car insurance market as example. Every car owner should know that {\it on average} he is paying more than he would get
back from the insurer. The expected cash-flow is toward the insurer, but the car owner can be nevertheless a happy payer (especially
he can get a contract from a competitive market). He can be happy even though he is aware of the fact that because of people like him
the insurer in general does get rich. The answer lies in the Von Newmann and Morgenstern's utility function. Since a consumer is generally risk averse, he would pay more than the fair expected value to rid himself of risks. A consumer gets income
from an outside economy, a sudden, rare and large  disruption to his wealth causes far more 
pain than the insurance premium he pays out. On the other hand the insurer can pool many insured together and in general with
a larger capital base. His utility function is not as concave as the one for the insured, for that amount of the contract.
Therefore even though the cash is strictly conserved, but both gain in utility. So such transactions can be beneficial to both
sides even though average cash-flow is in one direction only.

Our producers are just like the happy car owner. They participate economy outside the market, their
main worry is how to enhance their productivity using their special expertise. They need a liquid market to ensure that any need 
can be promptly met. They are in general not interested in getting rich exploiting the market's inefficiencies since Nature (in resources as well as in innovative ideas)
offers better yet opportunities for their expertise. Consider a producer who need to buy and sell in a market alternatively and
periodically. His action will make the market price more predictable, as a matter of fact very predictable if he acts periodically. In a thin market
with little speculator's participation he will have to suffer considerable market impact loss, in the applications section we
show such loss is proportional to $1/\sqrt v$, $v$ being the market volume per unit time. With the active participation of 
speculators, since his action is probably anticipated, a lot of willing speculators would like to be on the other side of the
transaction, making his impact on the price less pronounced.

We may view that the producers and the speculators live in symbiosis: Producers get main income from exploiting outside 
economy.   
He foregoes part of this income through market participation because that  he pays less attention than speculators to price 
inefficiencies. Speculators, on the other hand, having less ability to extract income from the outside economy, work hard to 
detect market inefficiencies. It is clear that more speculators around, especially those smart ones who can detect these
inefficiencies, make the market more efficient. In the next section we shall show that even the best speculators
cannot make these inefficiencies to disappear.

\section{Negative Entropy, a Measure of Market Inefficiencies}

If prices were a pure random walk, the variations would be a completely uncorrelated string of numbers. In physics and 
information theory\cite{shannon} we would say that that such a string of variables is completely disordered, or the entropy is 
maximized.
On the other hand, if the the price variations are somewhat correlated, then the entropy does not attain its maximal value.
It is convenient to consider actual values of the price variation's entropy with reference to its maximal value. Any difference
is therefore called negative entropy and it is taken as a measure of predictability. Shannon entropy is defined
as 
\beq
S(\mbox{\boldmath $i$})=-\sum_{\mbox{\boldmath $i$} }\rm{p}(\mbox{\boldmath $i$})\log \rm{p}(\mbox{\boldmath $i$})  
\eeq
Shannon entropy was introduced for information transmission, and later on it found uses in cryptography and other applied mathematical domains. It is strange that it did not find wide use in modern economics, considering the fact that a huge industry is engaged in
forecasting financial time series.

However, Shannon entropy is not yet the measure of predictability of a signal, it just says that how likely a particular
sequence would appear. For our purpose, we need to consider the so-called conditional entropy, a less known relative of the
Shannon entropy:
$$
H(\mbox{\boldmath $i$}\rm{})_j=-\sum_{\mbox{\boldmath $i$}\rm{}}p(\mbox{\boldmath $i$})\rm{}\sum_{j}p(\mbox{\boldmath $i$}\rm{}|j)\log p(\mbox{\boldmath $i$}|\rm{}j)
$$
where $p(\mbox{\boldmath $i$}|\rm{}j)$ is the conditional probability that with the event $\mbox{\boldmath $i$}$ given, the event $j$ would follow (e.g. in sect.2 $p(+,+,-|+)$, {\boldmath $i$}=$+,+,-$, j=+).         
Then the partial entropy has to be averaged over all the events $\mbox{\boldmath $i$}$. It is easy to show that
the conditional entropy can be expressed through the usual Shannon entropy:
$$
H(\mbox{\boldmath $i$}\rm{})_j=S(\mbox{\boldmath $i$},\rm{}j)-S(\mbox{\boldmath $i$}).
$$
The  conditional entropy has the meaning that with an previous event given, i.e. the past price, what can one 
predict the future price movement. If the conditional entropy is maximized, then it implies that the future event $j$
is completely unpredictable. Any departure from the maximal value represents the potential edge for
speculators, albeit only probabilistically.

The so-called negative entropy ($\Delta H$) can be defined as the difference between the maximal value and
the actual value of the conditional entropy $H$. For the binary case when $\boldmath i$ and $j$
are both binary strings, the maximal value is one. We have:
$$
\Delta H=1-H.
$$

\section{Persistent and Anti-Persistent Walks}

Persistent and Anti-Persistent Walks (PW and APW) were first introduced by Mandelbrot. But his definition
refers long range tendencies and the diffusion exponent are different from 1/2. Here we limit ourselves to short
range correlations, as it seems to be the case in the market prices. Many proponents of the Random Walk Hypothesis (RWH)
probably ignore the fact that being a random walk does not necessarily imply efficient markets. Short time correlations in prices can
still arise and profitable strategies can still be designed.

 Let us consider a walk $x_t$ which can go up and down one unit, in discrete time steps ($t=1,2,...$). 
The walk does not have preference (up or down)
but has the tendency to {\it continue} its last direction with probability $p$, to change direction with $1-p$. Such a walk
strictly speaking is not a Markovian process since its history is needed to know its future direction. But the long time   
correlation function decays exponentially $<x(t)x(0)>\sim\rho^t=\exp-t|\ln\rho|$, where $\rho=|2p-1|<1$. A walk with $p>1/2$ is said to be a PW since there is a probabilistic
bias to continue in the same direction; whereas $p<1/2$ implies an APW which tends to zigzag. $p=1/2$ corresponds a perfect RW.
It is clear that for times much larger
than $t^*=1/|\ln\rho|$ the  PW or APW cannot be distinguished from RW. Let us mention that for PW and APW we have also normal
distribution densities, for large times. If the corresponding RW has $p(x)\sim \exp(-x^2/2Dt)$, we now have $p(x)\sim \exp(-x^2/2D_pt)$ with
$D_p=p/(1-p)D$. We see that for $p$ large the diffusion constant is larger than the RW's $D$, meaning that the PW prices tend to fluctuate
wider than RW; for small $p$, we have  $D_p<D$, meaning that the APW prices fluctuate less.

Notice that in the previous section the equity index seems to be a PW, while the Forex and Commodity prices
resemble more an APW. Elsewhere we shall report that this is actually a generic tendency, that there are two large classes
of prices and they can be represented by either PW or by APW. Of course for real prices we need more than just one step 
histories to help predict the next move, but PW and APW offer the simplest examples of inefficiencies in market prices.

Using the above definition we can easily evaluate the conditional entropy for both PW and APW. It is equal to 
$H=-[p\log_2p+(1-p)\log_2(1-p)]$ and $\Delta H=1-H\approx \rho^2$, for $\rho=|1-2p|$ small. We obtain the answer in closed form here because PW and APW are dependent only on 
one-step histories. Using this simple example we can illustrate why the conditional entropy is useful.  

\section{Why Market Inefficiencies cannot be arbitraged away}
The standard assumption in the mainstream economics is that if there is an arbitrage opportunity,
"smart" investors would spot the chance and make a profit, thereby making the opportunity 
disappear in no time. It hardly occurs to people that in the competitive, fair markets the most
frequent profit opportunities are only {\it probabilistic} in nature. To profit from probabilistic
opportunities one has to bear risks. Moreover, the "smart" investor's capital is finite so must 
also be his impact on reducing the markets' inefficiencies. We shall see that even with the smartest 
investors, one still needs {\it infinite} capital to make such inefficiencies {\it completely} 
disappear. The standard literature is conspicuously vague on this point. Of course when the 
arbitrage opportunities are "risk-free", as emphasized in the literature, they 
could be arbitraged away.  

Therefore we are led to consider that a speculator is faced with a market and his tools can 
detect some probabilistic edge. In our representation, we can speak of the negative entropy 
being non-zero. Let us recall in our simplest version we have $\Delta H\sim \rho^2$, and $\rho>0$.
Elsewhere it has been shown that with a repeated investment game and a constant favorable chance
$\rho>0$ the most profitable (aggressive) strategy is the Kelly's method.\cite{Kelly} It can be 
shown\cite{mz} that no smart investors can do better than Kelly in the statistical sense, given the same edge.

According to Kelly's method, an investor should invest an optimal fraction of his current capital,
to achieve the most aggressive profit. It is for this most extremal case we want to show, that still
infinite capital is needed. One can easily determine Kelly's optimal fraction by considering the
"typical" value of the cumulative capital to be maximal. In the case at hand, denote the current
capital by $C$, the optimal fraction is $f^*=\rho C$\cite{mz}.

Let us assume that the current market conditions can sustain the marginal inefficiency $\rho$ and
with speculators' volume $v$ feeding on it, at a given frequency. Additional volume $\Delta v$ 
of the speculating capital of the same type would make the market inefficiency weaker ($\rho'=
\rho+\Delta \rho<\rho$). We can estimate the relationship. The additional volume $\Delta 
v$ would damage the existing edge, and now we have 
$$\rho'=\rho (1-\Delta v/v).$$
The RHS of the above equation has a factor $\rho$, since the additional capital can only "damage"
a fraction $\rho$ of the market price, $\Delta v/v$ is the relative strength. The above equation
can readily be integrated, to obtain the simple relation:
$$v\sim 1/\rho .$$
Recall that even the most aggressive speculator risks only a fraction of his capital, we thus 
have:
$$C\sim 1/\rho^\gamma ,  \gamma=2 .$$

This implies that the total engaged risk capital for the most aggressive speculators follows 
a power law, the simplest estimate yields the exponent to be two. For more general cases with
the negative entropy $H$, we have 
$$C\sim 1/H^{\gamma/2} .$$
Upon additional speculator's capital engaged, the market becomes more and more efficient, but
never completely efficient. To have the market really efficient $\rho=0$ it is clear infinite
risk capital is needed. 

Note that as the favorable edge $\rho$ diminishes, there is no reason why the speculator should
continue to pour in additional capital. Also note that the above estimate is only for the 
extremal case. It sets the lower bound for necessary risk capital or, equivalently, the upper bound
on the impact that a given amount of speculator's capital can do to. The above estimate is 
however not complete. We have considered only one given strategy and one frequency at which the inefficiency $\rho$ and
speculator's action operate. As for many nonlinear system, different frequencies are necessarily
coupled,  the Naviers-Stokes equation for turbulence is a typical example. In our analysis, we do not
have yet a complete model describing the interaction of different frequencies. But the analysis 
of a simplified market competition model offers hope\cite{czs}. So our estimate of
the power law exponent can only be qualitatively correct. We want to show that this exponent
depends on detailed interaction between the producers and speculators. In the following section 
we shall show that producers like to have a liquid market, typically they pay a cost of illiquidity
proportional to $1/\sqrt v$. This implies that the larger is the volume (per unit time), the 
smaller is the cost. It is plausible to expect that as $v$ increases, the producers would pay less
market impact costs thus they are more inclined to use the market. On the other hand, the 
aggregate gain by speculator's cash-flow $\Delta H v\sim \rho^2v$ must be provided by the producers. 
The simplest assumption would be that the total cash-flow is kept constant, since a more 
efficient market (smaller $\rho$) would induce the producers to use more often the market.
This leads to another scaling relation
$$
C\sim 1/\rho^\gamma , \gamma=3 .
$$
This example shows that the ultimate exponent $\gamma$ should depend on the precise type of
market interaction, but it is likely that the power law relation would stay for a broad spectrum
of markets.

In the following section we examine various direct and indirect implications of MEM.

\section{Intrinsic Illiquidity cost and Market Impact equation}
In this section we digress a bit from the main line of this work to consider some implications
of MEM to market dynamics. To understand what a small inefficiency can do to the market, first
we need to discuss what are the implications of an efficient market. In the literature\cite{hara} much detailed mechanism is discussed.
 
The central point of this work is that markets are almost efficient with the inefficiency margin
small. As a consequency any arbitrage opportunity without risk should be absent. Starting from
this general principle, we can derive some useful, simple conclusions.

1) {\bf Illiquidity cost}.
Market prices are usually quoted as bid and ask prices. The difference is called the spread.
Thus an investor can buy at the higher price (ask) and sell at the lower one (bid). The market 
maker would try to quote a competitive spread to attract both sellers and buyers, but he has to
make sure that himself is not {\it excessively} exposed to market risks. His spread contains
a margin to fend off risk, as well as a smaller part for his mediation work. We suppose that the
former is more important and intrinsic, since on this margin he cannot afford discount. 
This margin we want to call the Intrinsic Illiquidity Cost.
It is empirically observed that the more liquid is a market, the smaller is the spread. There
is extensive literature on the composition of this spread\cite{CLM}, however we would simplify
the problem so that a generic law appears.

Market transactions are seperated by some time lapse intervals, even for very liquid markets.
Econometricians use the term of non-syncronical trading. When a buy or sell market order comes to the
market maker (e.g. specialist in stock exchanges), he has to temporarily absorb the trade since
his job is to {\it make} the market. He takes the trade only because that he can  count on to give the trade
to somebody else, a short time after. The longer the waiting time, the more risk he bears of 
holding the bag. In order to protect himself from exposing to excessive risks, he posts 
two prices, bid and ask ($P_b, P_a$), i.e. buying low and selling high ($P_b<P_a$).

It is instructive to think that that $intrinsic$ prices are continuously moving as a RW, with
or without actual trading activities. This is because that real markets are exposed to all sorts
of outside and inherent noise factors, e.g. news, changing moods, other related markets etc. It is useful
to represent this continuous price as a price proxy. Now let us consider the total daily volume
to be $V$, expressed through a basic unit which cannot be further divided (like a SP500 futures
contract). Denote also by $v=V/T$ as the volume per unit time, $T$ being the daily trading duration. Clearly the typical waiting time $\tau$ between two consecutive trades is inversely
related to the volume, $\tau\sim 1/v$. On the other hand, during the interval $\tau$ one
should expect that the market price proxy nevertheless moves ahead as a RW, to the leading order approximation. So his spread should
cover the most probable fluctuation range $\pm \sqrt\tau$ of this price proxy. This leads us to the estimate
of the so-called intrinsic illiquidity cost 
$$
{\rm spread}\sim 1/\sqrt v.
$$
Of course this spread does not protect the market maker from risk with certainty, notably when
in the asymmetrical information situations\cite{hara}. However in most fair trading situations, the market 
maker should not charge much more than the above estimate, otherwise he can almost
make a certain gain since out-range fluctuations are rare. He still can make a profit in the 
probabilistical sense: his probable edge discussed in previous sections $\rho$ is somewhere
between zero and one, depending on the coefficient in front of the above expression, 
as well as detailed market making mechanism. Return to our discussion of marginally efficient markets, the modification to the above equation is small and such modifications 
maybe overwhelmed by other factors in market-making. Unlike for the speculators the small probabilistic edge does not play an important role for the market makers. One just wants to make sure that market makers do not rip off other participants with certainty, using the above equation as a criterium. 

Clearly a market maker has to fine-tune his spread to be low enough
to remain competitive, in the same time high enough to rid him of the excessive risks and to 
make a living. The spread should be time dependent since during a trading day the volume varies
significantly. The above estimate is only in the average sense. One can as well use time intervals
smaller than daily to reflect the time dependence of the spread.

2) {\bf Market Impact Equation}.

A related problem is the so-called market impact of large orders. Empirically, one observes that
large orders move the price more than small orders. For example a buy order of large size would
probably push the price upwards markably. This can be neatly expressed through a simple model\cite{bc}
of price movements $\Delta P$
$$
\Delta P = \lambda (B-S).
$$

The parameter $\lambda$ represents the "market depth", $B$ and $S$ stand for buyer and seller
initiated orders. Positive values of $B-S$ mean that more buyers are eager to buy than sellers
are willing to sell, negative values the opposite. Of course every buy order is exactly
offset by a sell order---the very definition of the market transaction. We distinguish buyer- or seller-initiated trades. For example, if \ a trade takes places at the bid price, we can identify it as a seller initiated
trade since the seller is eager to low his expection to meet the buyer's wish price.

This problem has been the subject of a recent careful experimental work on DAX futures market\cite{dax}.
The authors find that the relationship between $\Delta P$ and $B-S$ is highly nonlinear, 
especially for large orders. They found that for large orders the price movements are much 
smaller than the above linear equation would imply.

Let us use our present MEM theory to analyse the problem. A large order will typically make a
large deviation on the price history. Without major news breaking out, such abrupt movement usually
goes back in a short time. This presents an almost certain arbitrage opportunity. According to
MEM, only probabilistical opportunities are allowed existence. If an opportunity deemed
by the market participants unusually attractive, many speculators on the side-line would jump
in to the fore to grab some profit. Let us estimate what price impact actually a large order
has: A larger order will take longer time to complete. Let us assume the necessary time is
linearly proportional to the order's size (this proviso is subject to further evidence).
A large order is defined not necessarily by a single trade, but should be consistent of a string of trades. It should be understood to be the cumulative imbalance between buyer- and seller-initiated trades, using a given time interval as units.

Suppose a large order comes and it starts to move price in one of the two directions. Many
speculators on the sideline duly take notice of this movement. From their experience, they
expect that price would bounce back and a profit opportunity is in the making. If they all jump
into the fore by taking the opposite side of the trade, the price hardly moves in the supposed
direction and the speculators can hardly expect a meaningful gain. On the other hand, if they wait too long, a large deviation  results from the
large order, this represents too good a chance to escape attention of vigilant fellow speculators.
There is a natural compromise point: the price would be let to deviate in one direction but
not too much. During the time interval $\tau$ of completing this large trade, the (RW) market price would allow a deviation of $\Delta P\sim \sqrt\tau$ and 
$\tau\sim (B-S)$, without making such an event too attractive. Therefore we have the following equation
$$
\dot P=\lambda_{BS}\sqrt {|B-S|}+\eta(t) .
$$
We have used a continuum notation, but this equation should be understood in discrete unit.
$\lambda_{BS}=\lambda_B>0$ if $B-S>0$, $\lambda_{BS}=-\lambda_S<0$ if $B-S<0$. Empirically
often $\lambda_B <\lambda_S$, i.e. a sell order's impact on price is larger than that of a buy order of 
the same size. This is because that stock markets have a upward bias, elsewhere we will study this difference in more details. The stochastic noise term $\eta(t)$ is added representing that the price is a RW,
in the absence of large orders.
The above equation is qualitatively in agreement with the experimental observation on the DAX
futures price movements. However, in view of the present proposal, more systematical verifications are  needed.

\section{Summary and Future Work}
 
In the above discussion we have tried to outline a novel approach to study dynamics of markets.
The key ingredient is to recognise the important role played by the marginal inefficiency of the
markets. We are led to view the market economies as a web of agents, quite like the food chain in ecology, connected through the markets. Two groups of agents (producers and speculators) live
in symbiosis, the one injects "negative entropy" to make markets attractive; the other tries to
exploit the inefficiencies thus providing a social service without wanting. These two groups 
can be identified on each level and nodes of the economy-web. The toprisks to enjoy some fruits of Nature. Thus this web normally absorbs the original inevitable (since Nature's offering is never certain!) noise shocks, through different layers. At the bottom are people who do not even
bother to think about investments, they just put their extra wealth in bank or worse in cash. Surely they shall feel the least shock effects but also miss out the oppurtunities of Nature's
generous offering. Elsewhere\cite{zhang} we shall show that such a web is capable of nasty surprises: the original inevitable shocks are in general amplified, due the over- and under-reaction through each layer of the web. Here again lies the imperfect market mechanism,
which results in extra or excessive fluctuations that Shiller\cite{shiller} has studied.

{Acknowledgements: during the past few years I have benefitted from fruitful collaborations with D. Challet, G. Caldarelli, M. Marsili, and F. Slanina}

{\bf Figure Captions:}

Fig.1 Daily price closes of NY 400 Composite Index during 32 years. Using this data shows the conditional
probabilities are plotted. Conditional on the past observations (e.g. +++), this figure 
shows the probability to have the next day price variation to be +. The dots in a columne represent the relative importance of sampling.

Fig.2. The data in Fig.1 now is broken into 4 segments, then the conditional probabilities are
calculated for each of them. We can see from each period of eight years to the next, there is some overlap. This shows that market "habits" are somewaht persistent, allowing profitable strategies to be designed.

Fig.3. Similar to Fig.1 but for high frequency data (every 1/2 hour) of Forex (DM and SF) and Commodity (Silver) prices. The interval is not fixed, but the price variations have to be larger
than a threshold value to be considered. Especially if we see the Silver price have two consecutive --, then there is almost 80\% probability to be followed by a +. Probabilistic edge
this good are exception and deceiving. One has to consider the Silver price has very violent
swings with tiny volumes, making real speculation impractical. But in reality the edge as small as 55\% can be profitable.

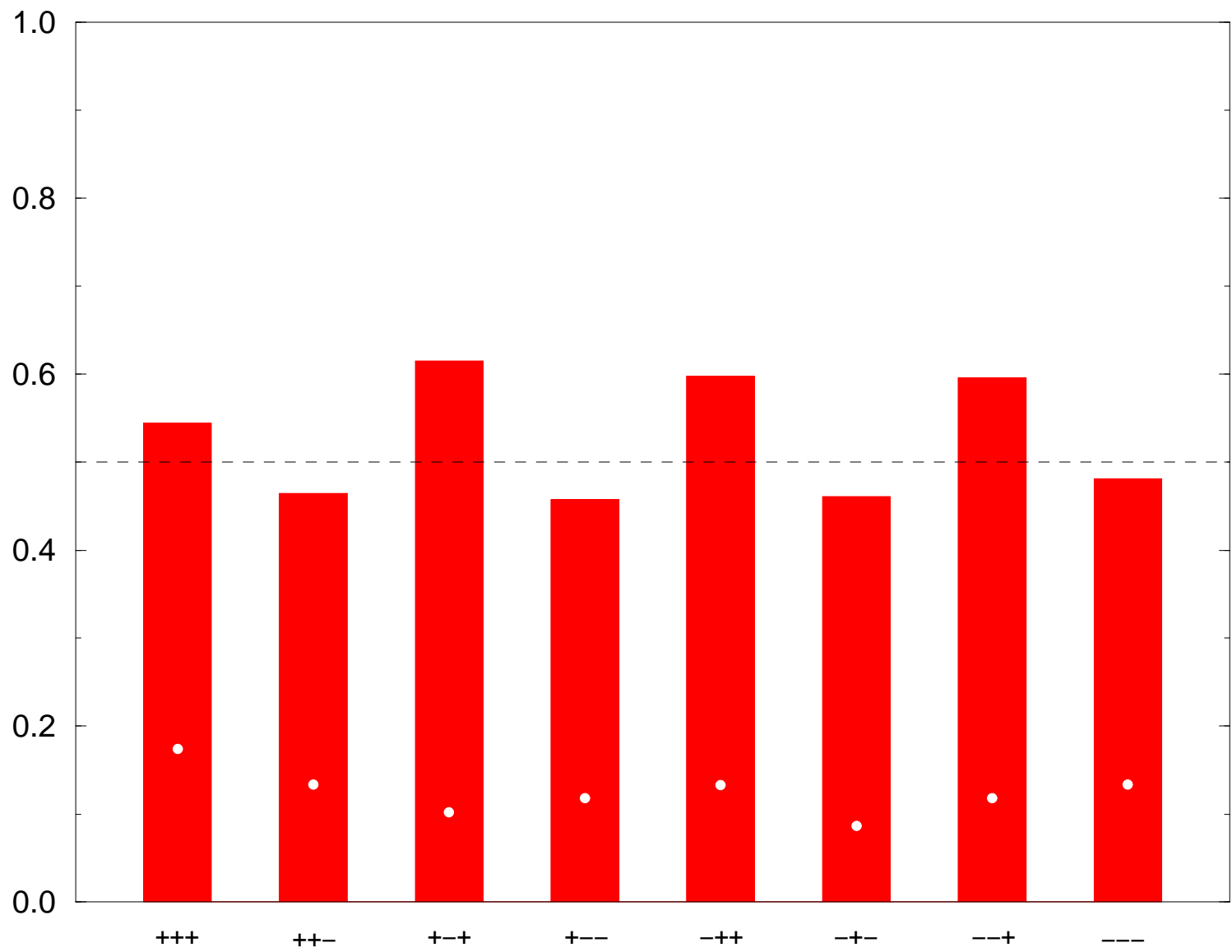
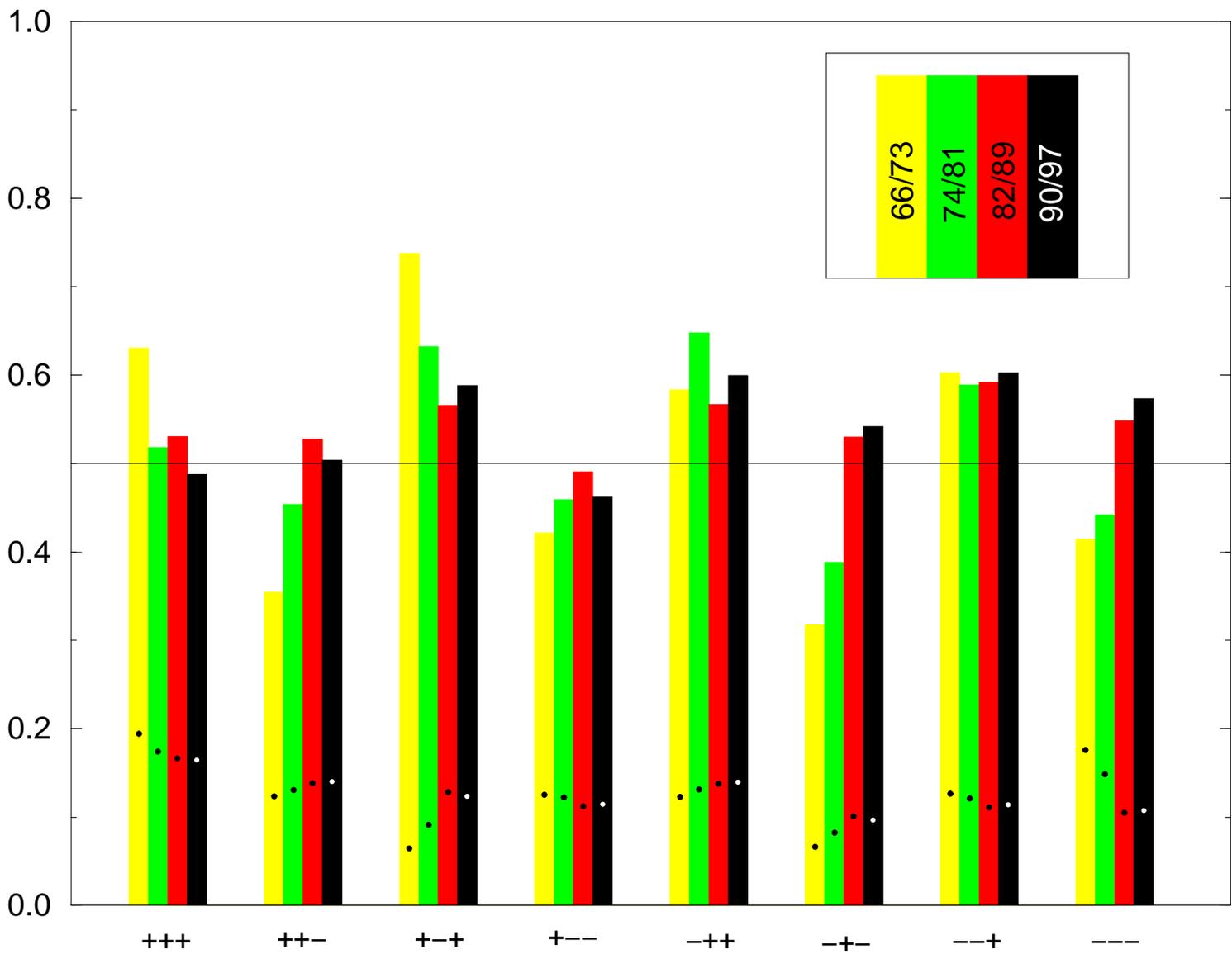
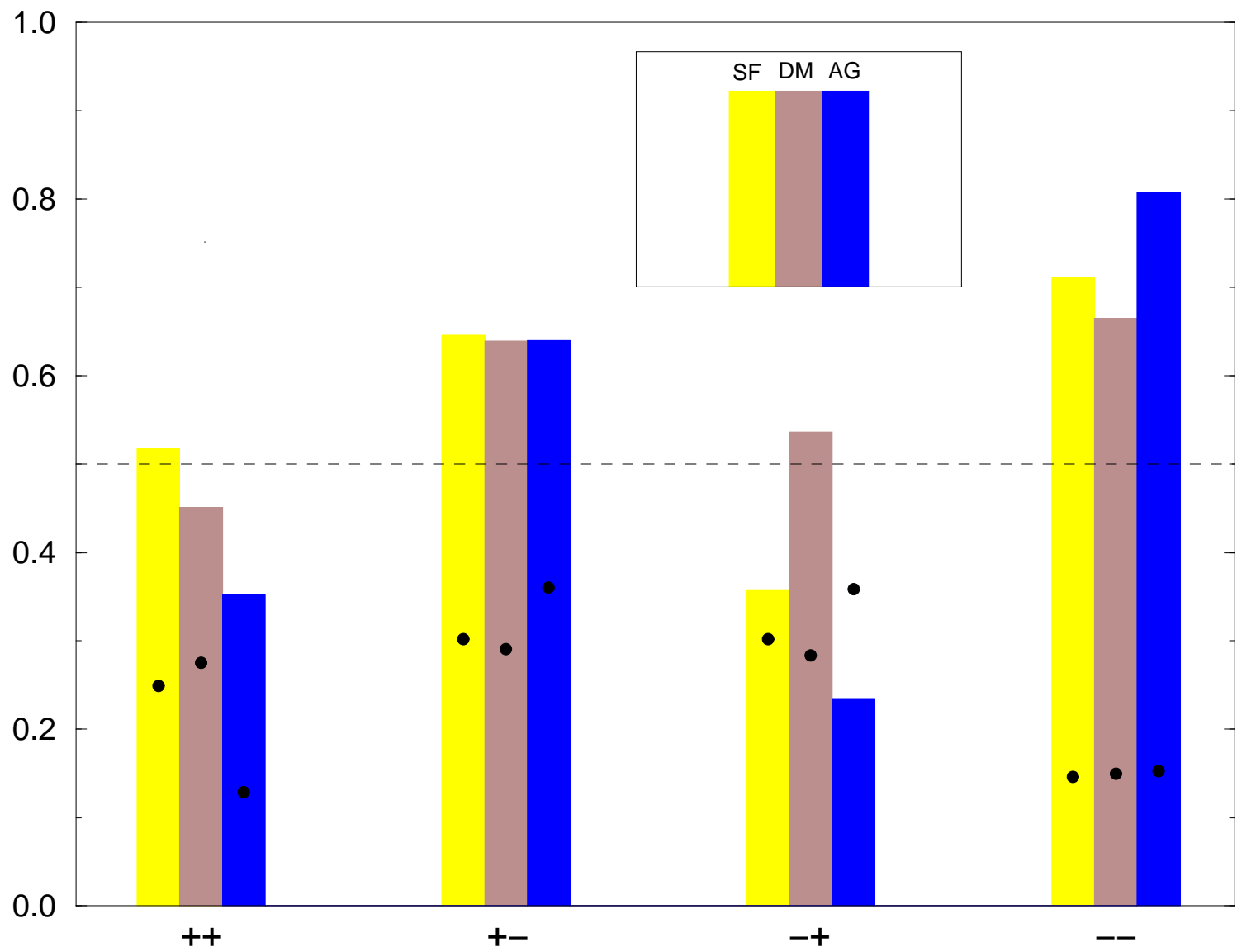
\end{document}